\documentstyle[twocolumn,aps,prl]{revtex}
\include{epsf}
\epsfverbosetrue
\begin{document}
\twocolumn[\hsize\textwidth\columnwidth\hsize\csname@twocolumnfalse\endcsname

\draft

\title{Ge-substitutional defects and the
$\sqrt{3}\times\sqrt{3}$ $\leftrightarrow$
3$\times$3 transition in $\alpha$--Sn/Ge(111)}
\author{Jos\'e Ortega$^{(1)}$, Rub\'en P\'erez$^{(1)}$, Leszek
Jurczyszyn$^{(2)}$ and  Fernando Flores$^{(1)}$}
\address{$^{1}$
Departamento de F\'{\i}sica Te\'orica de la Materia Condensada, 
Universidad Aut\'onoma de Madrid, E-28049 Madrid, Spain}
\address{$^{2}$
Institute of Experimental Physics, University of Wroclaw,
Pl.~Maksa Borna 9, 50--204 Wroclaw, Poland}
\date{submitted \today}
\maketitle

\begin{abstract}
The structure and energetics of Ge substitutional defects on the 
$\alpha$-Sn/Ge(111) surface are analyzed using Density Functional 
Theory (DFT) molecular dynamics (MD) simulations. 
An isolated Ge defect induces      
a very local distortion of the 3$\times$3 reconstruction,
confined to a significant downwards displacement 
(-0.31 \AA) at the defect site and a modest upward displacement (0.05
\AA) of the three Sn nearest neighbours with partially occupied dangling
bonds. 
Dynamical fluctuations between the two
degenerate ground states 
yield the six-fold symmetry observed around a defect in the experiments 
at room temperature.
Defect-defect interactions are controlled by the energetics of the
deformation of the 3$\times$3 structure: 
They are negligible
for defects on the honeycomb lattice and quite large for a third defect
on the hexagonal lattice, explaining the low temperature defect ordering.
\end{abstract}


]

\narrowtext

The driving force behind the T-induced reversible 
$\sqrt{3}\times\sqrt{3} \leftrightarrow 3\times3$ phase
transition observed for 1/3-ML coverage of Sn or Pb on Ge(111) 
has been under intense debate since its discovery 
in Pb/Ge(111)\cite{Carpinelli96}.
After the early suggestions of surface Fermi wavevector nesting or electron
correlations, leading to
the formation of a Charge Density Wave \cite{Carpinelli96,Carpinelli97} 
at low T, recent theoretical work
indicates that the softening of a 3$\times$3 surface 
phonon\cite{Perez01}, associated with the electronic energy
gain due to the Sn dangling bond (DB) rehybridization, 
plays a major role on this transition. 
The dynamical fluctuations model\cite{Avila99} associated with this soft 
mode provides a consistent account of the main
experimental facts, in particular the similarities and differences between the
isovalent Sn/Ge and Sn/Si(111) systems\cite{Perez01}.  

DFT calculations for the
$\alpha$--Sn/Ge(111) surface predict that the ground state 
is a 3$\times$3 reconstruction in which
1/3 of the Sn adatoms are placed $\sim$ 0.3 \AA\ higher than the other
Sn adatoms
\cite{Perez01,Avila99,Ortega00,Gironcoli00}.
In other words, out of the three
3$\times$3 sublattices that form the $\sqrt{3}\times\sqrt{3}$ lattice
\cite{Melechko00}, ``up" Sn adatoms ($Sn_u$) occupy one of these
sublattices (say $a$),
forming a 3$\times$3 {\em hexagonal} lattice, while sublattices $b,c$
are occupied by `down' Sn adatoms ($Sn_d$) (forming a {\it honeycomb} pattern).
These calculations also show that this reconstruction is correlated with
a change in the surface electronic structure\cite{Ortega00}:
the $Sn_u$--DB is
now fully occupied while the other two $Sn_d$--DBs share the remaining
electron per 3$\times$3 unit cell.
STM experiments are naturally explained in terms of these different DB 
occupancies.
At LT, below $T_c$, filled-state STM images show 
the 3$\times$3 phase : bright spots occupy the 
3$\times$3 sublattice associated with the $Sn_u$--DBs,
while the other two sublattices, associated with the $Sn_d$--DBs, 
are occupied by darker spots\cite{Jurczyszyn01}.
At RT,however, dynamical fluctuations of the Sn adatom
heights, fast with respect to the STM scan speed,
induce a $\sqrt{3}\times\sqrt{3}$ periodicity in the STM images. 

While these models assume an ideal 1/3-ML Sn surface coverage, 
point defects, with an average concentration of 3$\pm$1 \%
\cite{Melechko00}, are always present.
90 \% of these
defects are Ge substitutionals (called Ge defects in the rest of the paper), 
while vacancies account for the other 10 \%.
STM, a real-space experimental technique, has been the method of choice to get 
non-averaged information 
on the structure and distribution of these 
defects on Sn/Ge(111)\cite{Melechko99,Weitering99,Melechko00}
across a wide temperature range,
covering the transition temperature $T_c$. 
Three are the main experimental results:
(i)  At RT Ge defects are imaged as dark spots in the
$\sqrt{3}\times\sqrt{3}$ lattice 
 surrounded by
six features (brighter than Sn adatoms away from defects) that
correspond to the six nearest-neighbours (n.n.) Sn atoms.                       
(ii) At LT most of the Ge defects occupy positions corresponding
to the two honeycomb $Sn_d$--sublattices $b,c$, with equal probability,
and are therefore surrounded
by three bright and three darker Sn n.n. 
(iii) At RT Ge defects are randomly
distributed over the three sublattices $a,b,c$.

These  results have been considered as supporting evidence for an
interpretation of the $\sqrt{3}\times\sqrt{3} \leftrightarrow$ 3$\times$3
phase transition in terms of the interaction of the Ge 
defects\cite{Melechko99,Weitering99,Melechko00}.
One of the main ingredients in this argument is the six-fold symmetry
assumed for the perturbation induced by an isolated defect.
This six-fold symmetry (honeycomb, with one atom dark and two atoms bright
in the three possible sublattices), suggested by the RT STM images, 
is, however, in contradiction with the
hexagonal pattern (two atoms dark and one atom bright, with three-fold
symmetry) observed in the filled-state images below $T_c$.
In order to explain the LT experimental results, 
a defect-defect interaction, with a temperature dependent decay length
$l(T)$ (e.g. $l(300 K) \simeq $ 11 \AA, $l(100 K) \simeq$ 100 \AA) 
has to be introduced. This long--range interaction
would be responsible for the ordering of defects at LT on two of the
three possible sublattices. Finally, the superposition of the
perturbations induced by defects in these two sublattices would give rise
to the 3$\times$3 phase.

In this letter, we show that  {\em ab initio} DFT-LDA  
MD simulations of  Ge substitutional defects 
on the Sn/Ge(111) surface support a completely different point of
view: Both the structure and the ordering of the Ge defects at LT are
induced by the ground state 3$\times$3 reconstruction. 
In particular, we have studied the atomic
geometry, the electronic structure
and the energetics of one and two Ge defects in a large
9$\times$9 unit cell. In the case of two defects, we have considered
three cases, corresponding to
the defects on first, second and third n.n. positions.
Our analysis shows that Ge defects at LT always create a three-fold
distortion around them, the simulated STM filled-states image showing
three bright and three darker spots on the n.n. Sn atoms.
The distortion is very local, affecting only to the Sn atoms which are 
nearest neighbours of the defects.
We have also performed first-principles MD simulations
of a 3$\times$3 unit cell with one Ge defect.
Our MD simulations indicate that the six-fold pattern observed around
Ge defects in the RT STM images should be understood in
terms of the RT dynamical fluctuations of the six n.n. Sn adatoms.
These dynamical effects contribute, together with
structural and electronic effects,  
to the enhanced brightness of the Sn atoms around the defect.
    

In our approach we calculate different defects using a
Sn/Ge(111)-9$\times$9 unit cell with a slab of 4 Ge layers and a layer
of H atoms saturating the dangling bonds of the deeper Ge layer (432
atoms in the unit cell). Both
the H layer and the last Ge layer are kept fixed during the calculations
to mimic  the bulk continuation. In our calculations we have used an
efficient MD technique, the Fireball96 code
\cite{Fireball96}, that allows us to analyze the large 9$\times$9 unit
cell we have considered. 
We have checked that this code
yields accurate surface geometries by comparing it with a
plane wave code (CASTEP)\cite{castep} for the case of the Sn/Ge(111)-3$\times$3 
reconstruction\cite{Ortega00} and for the case of one Ge defect on a 3$\times$3 
surface unit cell (see Table I).
 These results confirm the quality of the calculations
provided by the Fireball96 code and justify the wide use we are going to
make of it in the rest of the paper.
STM images are calculated from the {\em ab-initio} electronic structure
of the surface using the Keldysh-Green function formalism and considering
a W tip (for details on the method and the application to the
Sn/Ge(111)-3$\times$3 system see ref. \cite{Jurczyszyn01}).

\begin{figure}[htbp]

\hspace*{+0.00cm} \epsfxsize=8cm \epsfbox{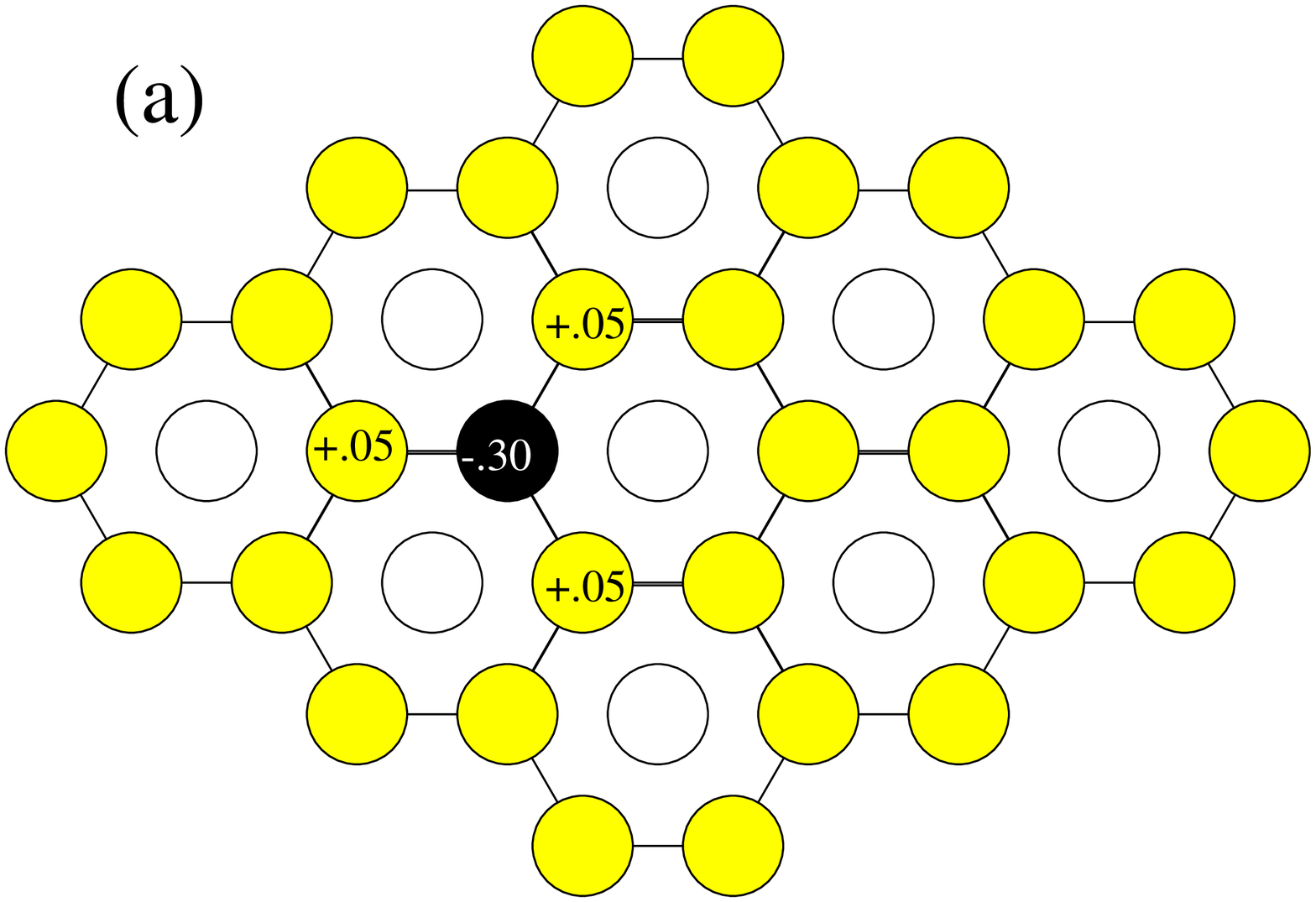}

\hspace*{+0.00cm} \epsfxsize=8cm \epsfbox{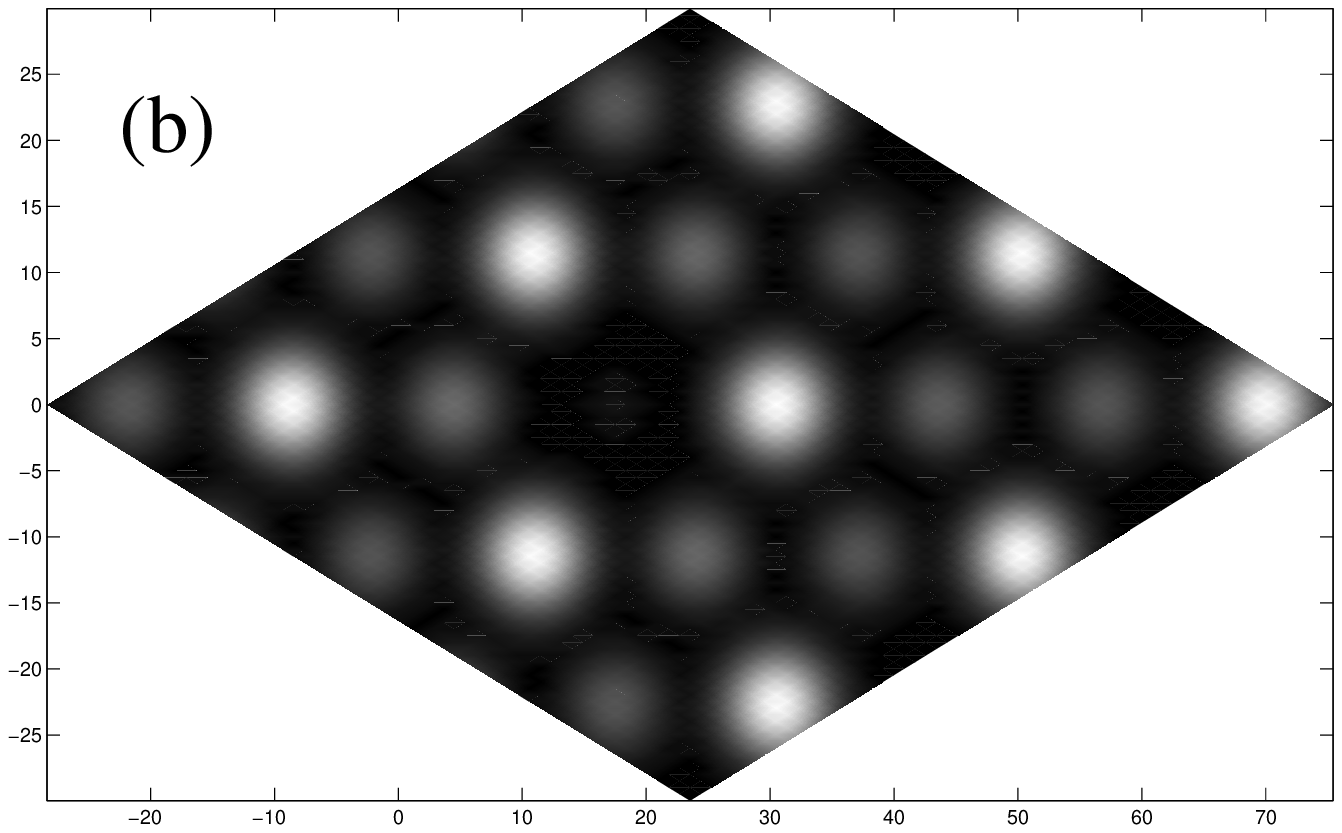}

\caption{(a) Relaxed structure for the isolated defect. 
Only the vertical displacements (in \AA) with respect to the ideal 
3$\times$3 structure which are larger than 0.01 \AA\ are indicated.
Notice that the distortion is confined to the defect site (black circle)
and the three neighbouring Sn atoms with partially occupied dangling bonds
(gray circles). Sn atoms with fully occupied DBs are indicated by white
circles.
(b) Filled-state STM image for the isolated defect. The bright protrusions
correspond to the fully occupied Sn  DBs, and the gray ones to the Sn
atoms with partially occupied DBs. The image reflects the three-fold
symmetry of the ground state structure of the defect.
}

\label{fig:1d}
\end{figure}

Figure 1a shows the relaxed geometry obtained in our calculations
for the $\alpha$--Sn/Ge(111) surface with one Ge defect per 9$\times$9
unit cell (3.7 \% concentration). The surface presents the 3$\times$3    
reconstruction with the Ge adatom on one of the two
 3$\times$3--sublattices
(say $b$) that form the honeycomb pattern associated with $Sn_d$
adatoms.
Our simulations yield for
the Ge substitutional a downwards displacement of 0.30 \AA\ with
respect to the original $Sn_d$ position.
The perturbation induced by the Ge defect is well-localized, affecting
only to the three n.n. $Sn_d$ adatoms ($c$ sublattice).
These adatoms are displaced upwards $\sim$ 0.05 \AA \ due to the
transfer of charge from the Ge defect: the Ge--DB is empty
and each n.n. $Sn_d$--DB gains $\sim$ 1/6 electrons. The occupancy 
of the n.n. $Sn_u$--DBs
($a$ sublattice) does not change (they are already fully occupied),
 and therefore their position is practically the same as in the
3$\times$3 reconstruction free of defects.
Figure~\ref{fig:1d}b shows the calculated filled-state STM image for 
this geometry.
Probably the most remarkable feature is the three-fold symmetry of the
image around the defect (only three of the surrounding Sn atoms are
imaged as protrusions), in contrast with the six-fold symmetry assumed
for the perturbation induced by the defect from the experimental RT STM images.

Notice that the solution shown in fig.~\ref{fig:1d} is two-fold
degenerate: an equivalent solution  is obtained applying a symmetry 
operation that exchanges the two sublattices $(a,c)$ occupied by the 
up and down Sn.
This degeneracy will be important when discussing the effect of
temperature on the surface geometry and its corresponding STM image.

\begin{figure}[htbp]

\vspace*{-0.50cm}

\hspace*{+0.00cm} \epsfxsize=8cm \epsfbox{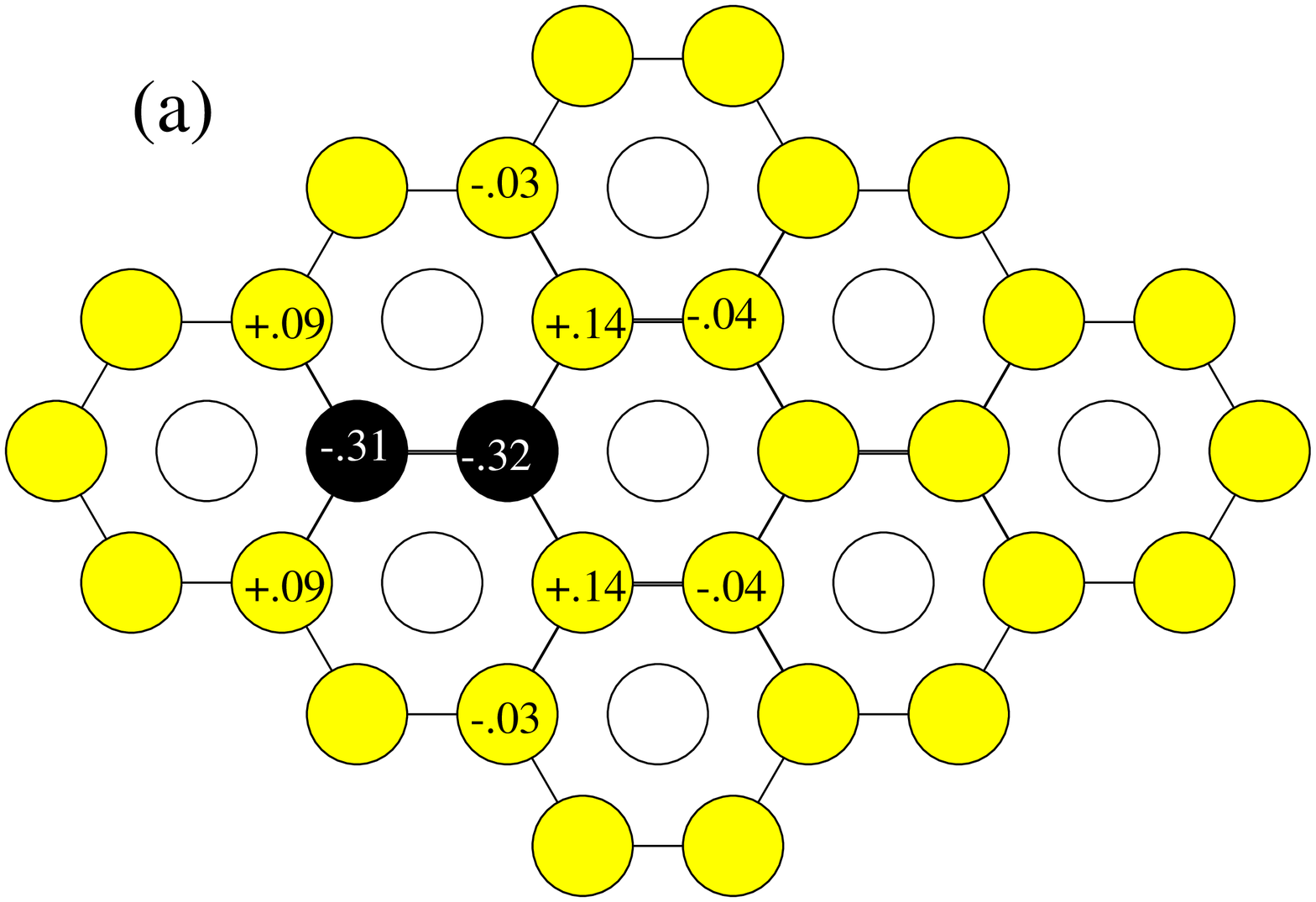}

\vspace*{0.25cm}

\hspace*{+0.00cm} \epsfxsize=8cm \epsfbox{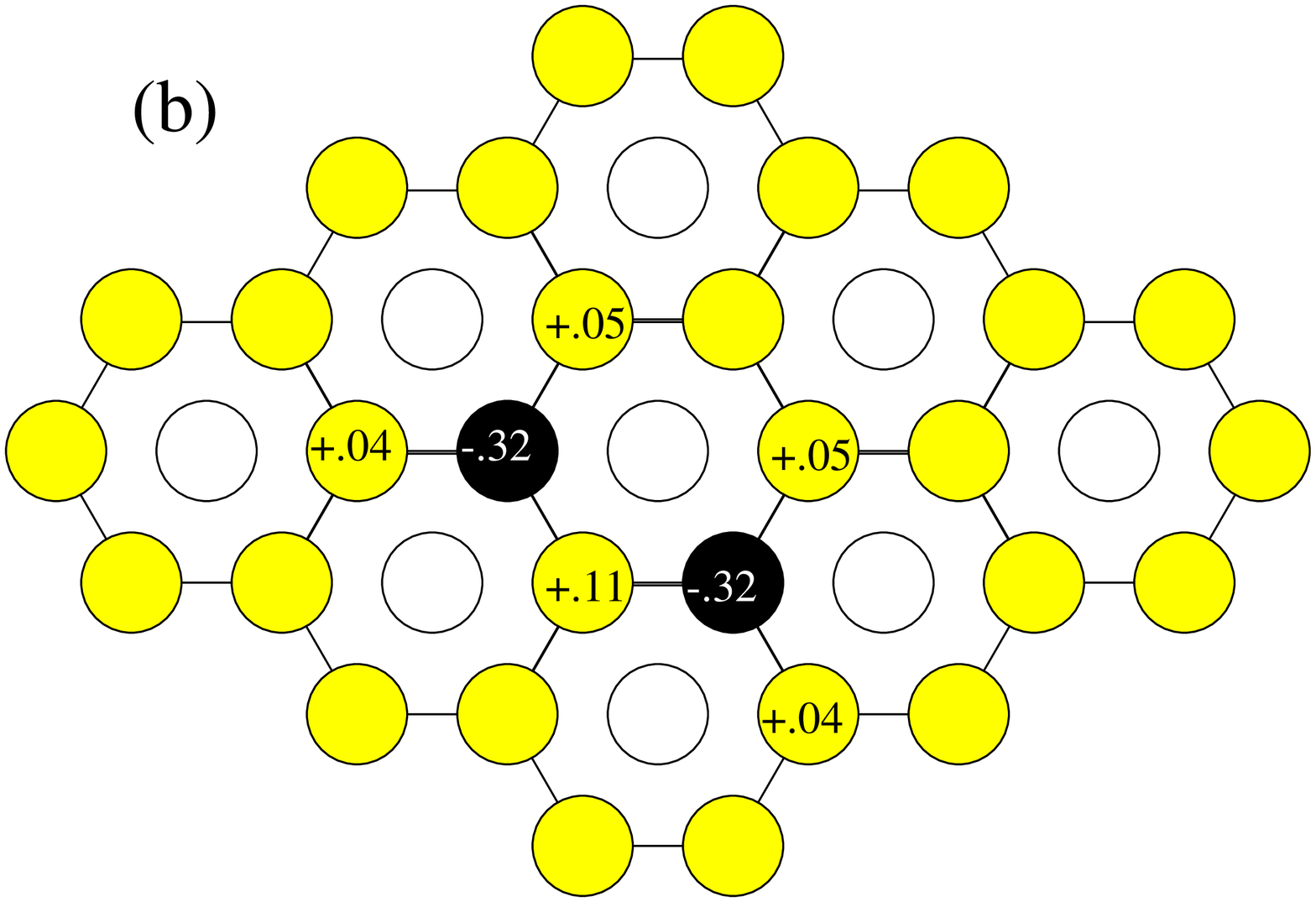}

\vspace*{0.25cm}

\hspace*{+0.00cm} \epsfxsize=8cm \epsfbox{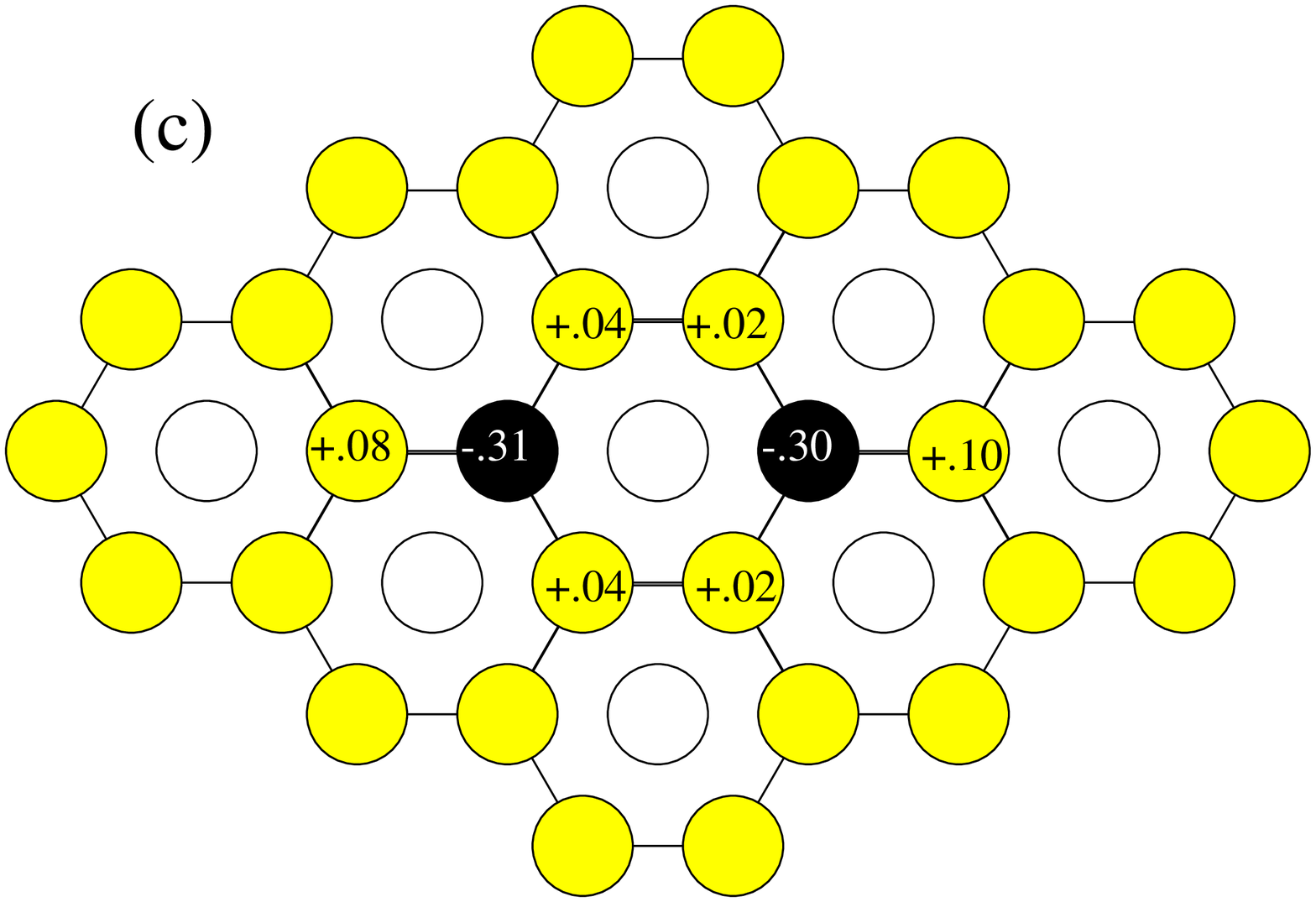}

\caption{Relaxed structures (as in fig. \ref{fig:1d})
for two Ge defects on first (a), second (b)
and third (c) nearest neighbour positions.
}

\label{fig:2d}
\end{figure}

Figures 2a,b,c display the relaxed adatom positions we have obtained
for the $\alpha$--Sn/Ge(111)--9$\times$9 surface with two defects on
first, second and third n.n. positions. In all the cases
the 3$\times$3 reconstruction adapts its particular geometry 
(out of a three-fold degenerate case) to the defect positions, 
in such a way that the two defects
are placed on the two $Sn_d$--sublattices forming a honeycomb pattern.
When the Ge defects are second n.n.
(fig. 2b) they are on the same sublattice (say $b$), and the ground state
is two-fold degenerate: 
an exchange of the $a$ and $c$ sublattices gives a completely 
equivalent solution.
In fig. 2a and 2c the Ge defects
occupy different sublattices and the solution is non-degenerate. 
As in the case of an isolated defect,
the perturbation induced by the defects is localized
essentially on the Sn-adatoms n.n. to the defects. 
The total energy for the three cases shown in fig. \ref{fig:2d} is,
within the accuracy of these calculations ($\sim 20$ meV/9$\times$9
unit cell), the same, confirming the localized nature of the distortion
induced by the  Ge defects.
It is remarkable that the  sum of the
upward displacements of all the neighbouring Sn-adatoms 
is, in the three cases, very similar to the sum of the displacements
induced by two independent defects ($\sim$ 0.30 \AA, see fig.
\ref{fig:1d}a).
This result suggests a roughly linear relationship between
the upwards displacement of a Sn adatom and the transfer of charge to its DB.

\begin{figure}[htbp]

\hspace*{+0.00cm} \epsfxsize=8cm \epsfbox{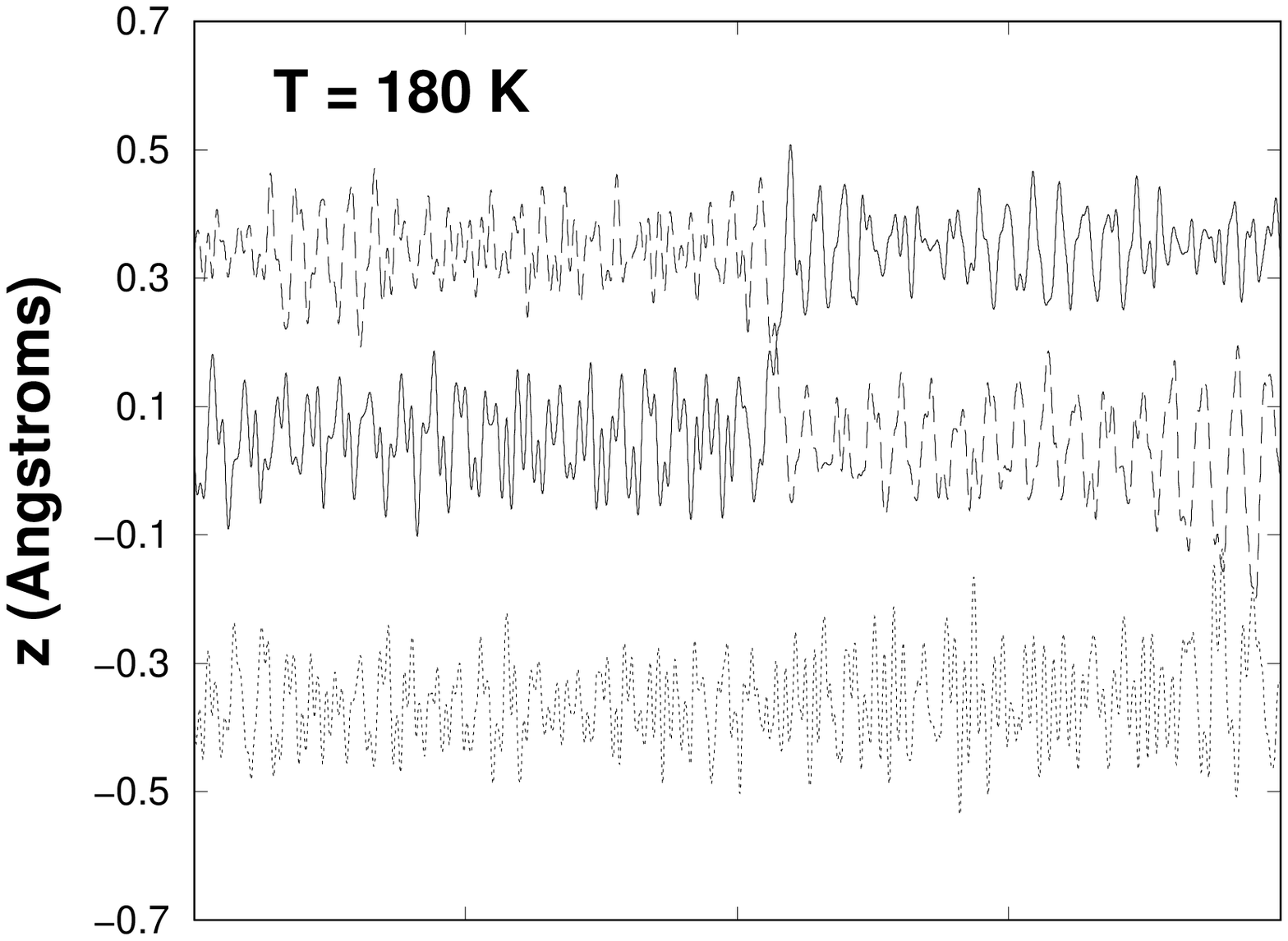}

\vspace*{-0.75cm}

\hspace*{+0.00cm} \epsfxsize=8cm \epsfbox{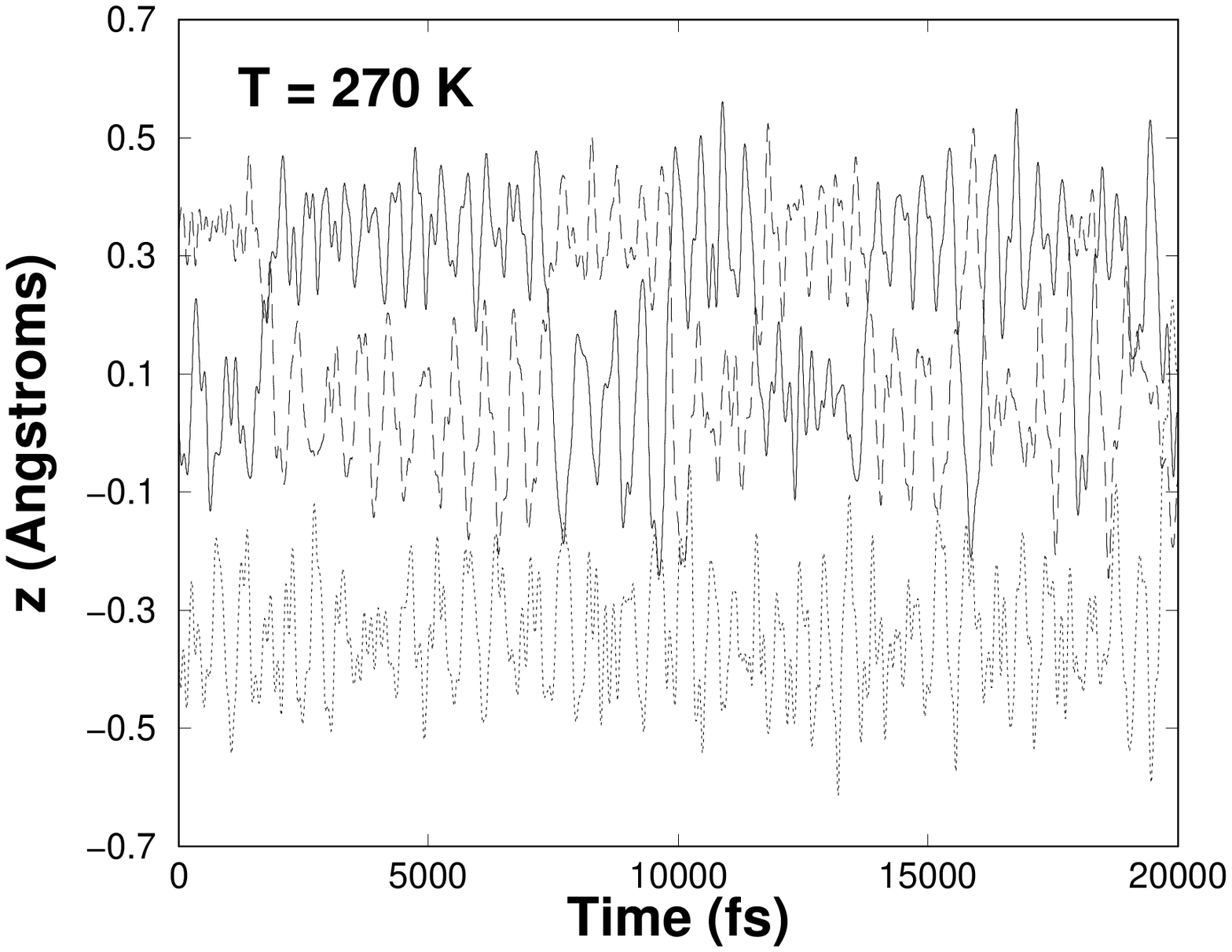}

\caption{ 
Time-evolution of the $z$-coordinate of the surface atoms (two Sn atoms
and the Ge susbtitutional --dotted line--) on the 3$\times$3 unit cell in MD
simulations at 180 and 270 K.
}

\label{fig:defect-md}
\end{figure}

Coming back to the experimental STM data for the Ge defects on the
Sn/Ge(111) surface, we first notice that our solution for a single
defect (fig,~\ref{fig:1d}b) has a three-fold and not a six-fold
symmetry, with the Sn atoms around the defect presenting  two different
types of displacements perpendicular to the surface.This apparent
contradiction with the experimental data at RT can be resolved, however,
considering the dynamics of the atoms around the defect. To this end, we
have considered a 3$\times$3 unit cell with a single substitutional  and
performed MD simulations at two different temperatures, 
T= 180 K and T= 270 K. 
The initial conditions
correspond to the positions of the relaxed structure of the defect (see
Table I) and random velocities following a Maxwell-Boltzmann
distribution. Then we follow the time evolution of the system
integrating the equations of motion with the forces obtained from the
DFT-LDA calculations. Figure \ref{fig:defect-md} shows the time
evolution of the $z$-coordinate of the three upper atoms 
(two Sn atoms and the Ge susbstitutional). These results show that Ge
remains around its equilibrium site, while the two Sn atoms tend to
exchange heights, very much in similarity with what we found for the
$\sqrt{3}\times\sqrt{3}$ structure~\cite{Avila99}.
Therefore, the RT STM images of the six Sn--adatoms n.n. to a Ge defect
must represent a 50 \% average of $Sn_u$ and $Sn_d$ (displaced upwards
$\sim$ 0.05 \AA \ w.r.t. normal $Sn_d$--adatoms,
 and containing 1/6 electrons more), while the RT STM
images of Sn--adatoms away from defects represent a 1/3 $Sn_u$ + 2/3 $Sn_d$
average. This dynamical (plus charge transfer) effect explains
in a simple way the different pattern observed at RT and LT on the STM
images around
a Ge defect (six-fold vs. three-fold symmetry) and also explains
why
Sn--adatoms n.n. to a Ge defect appear on the RT STM images
brighter than Sn--adatoms in regions free of defects.

We finally address the following question:
why do Ge defects at LT occupy preferentially
two 3$\times$3--sublattices, with equal probability,
and not the three of them or only one?.
Our results show that, on the $\alpha$--Sn/Ge(111) surface, 
Ge substitutional defects transfer their DB charge to neighbouring
Sn-adatoms DBs and do not participate in the up/down
dynamical fluctuations of the Sn--adatoms at RT.
Our results also show that at LT, when these fluctuations are supressed,
the stable configuration of the surface is a 3$\times$3 reconstruction
in which Ge defects occupy $Sn_d$ positions;
the defect-defect interaction is negligible
(as long as they stay on $Sn_d$ sublattices) and  Ge defects
occupy with equal probability any of the two $Sn_d$ sublattices.
A third Ge defect initially located in the other sublattice (hexagonal)  
should be forced, when the temperature is lowered, either to
present an upward displacement (contrary to what it does when it
is isolated) or to deform dramatically its environment in order to move
downwards. Any of these
solutions would imply a significant energy cost (160 meV for the defect
in an upper position according to our calculations), 
that the system prefers not to pay, locking in all the substitutionals 
in a honeycomb lattice. We conclude that the ordering of Ge defects
at LT is induced by the ground state 3$\times$3 
reconstruction and not the other way
around: Ge substitutional defects are not the driving force 
of the $\sqrt{3}\times\sqrt{3}$ $\leftrightarrow$
3$\times$3 transition in the $\alpha$--Sn/Ge(111) surface.

In conclusion, our DFT-LDA total-energy and MD calculations show that 
the ground state geometry around  a Ge substitutional defect corresponds
to a very local distortion of the 3$\times$3 reconstruction of the defect-free
surface, confined to the defect site and the Sn
nearest neighbours.
The lattice deformation can be characterized by a downwards
displacement of the Ge defect and upward displacement of the neighbouring
Sn atoms with partially occupied DBs, with respect to the ideal 3$\times$3
structure. This structural effect is correlated with the charge transfer
from the defect DB to the DBs of these Sn atoms.
This ground state configuration provides, at LT, an STM image with a three-fold 
symmetry, while at high T, the dynamical fluctuations between the two
degenerate states (with different Sn atoms in the upper positions) 
would yield the six-fold symmetry observed in the experiments.
These dynamical effects, together with the upward displacement and the
increased DB occupancy, are responsible of the enhanced brightness of the Sn
atoms around the defect.
Finally, the ordering of Ge defects at LT is naturally explained as an
effect of the 3$\times$3 reconstruction. 

This work has been partly funded by the spanish CICYT under contract
No.PB-97-0028. 
Part of the calculations have been performed in the CCC-UAM. 



\begin{table}

\vspace*{-0.50cm}

\caption{Atomic vertical displacements of the adatoms (in \AA), measured 
with respect to the 3$\times$3 reconstruction, for a Ge defect in a
3$\times$3 unit cell. 
For the Ge defect the displacement is referred to
the position of one of the ``down" Sn atoms of the 3$\times$3
reconstruction.
}

\begin{tabular}{dddd}
\ & Sn ``up" & Sn ``down"  & Ge defect \\
\hline
PW  & +0.04 & +0.08 & -0.31  \\
F96 & +0.02 & +0.10 & -0.32  \\
\end{tabular}                                                                   

\label{tab:r3to3x3}
\end{table}

\end{document}